\documentclass[]{IEEEtran}
\usepackage{lineno,hyperref}
\usepackage{amsmath,amssymb}

\newtheorem{defi}{Definition}

\usepackage{booktabs}
\usepackage{makecell}
\usepackage{multirow}
\usepackage{graphicx}
\usepackage{multirow}
\usepackage{tabu,bm}
\usepackage{color}
\usepackage{subfigure}
\usepackage{lipsum,mwe,cuted}
\usepackage{booktabs}
\usepackage{stfloats}
\hypersetup{hidelinks}

\usepackage{algorithm}  
\usepackage{algorithmic}
\usepackage{url}
\newtheorem{theorem}{Theorem}
\usepackage{cite}

\begin{document}
\title{Deep Learning Enabled Semantic Communication Systems}

\author{Huiqiang Xie, 
Zhijin Qin,~\IEEEmembership{Member,~IEEE,} 
Geoffrey Ye Li,~\IEEEmembership{Fellow,~IEEE,}\\
and Biing-Hwang Juang~\IEEEmembership{Life Fellow,~IEEE}% <-this % stops a space
\thanks{Huiqiang Xie and Zhijin Qin are with the School of Electronic Engineering and Computer Science, Queen Mary University of London, London E1 4NS, UK (e-mail: h.xie@qmul.ac.uk, z.qin@qmul.ac.uk).}
\thanks{Geoffrey Ye Li is with School of Electrical and Electronic Engineering, Imperial College London, London SW7 2AZ, UK (e-mail: geoffrey.li@imperial.ac.uk).
}
\thanks{Biing-Hwang Juang is with School of Electrical and Computer Engineering, Georgia Institute of Technology, Atlanta, GA 30332 USA (e-mail: juang@ece.gatech.edu).
}
}

\markboth{ }%
{Shell \MakeLowercase{\textit{et al.}}: Bare Demo of IEEEtran.cls for IEEE Journals}

\maketitle

\begin{abstract}
Recently, deep learned enabled end-to-end (E2E) communication systems have been developed to merge all physical layer blocks in the traditional communication systems, which make joint transceiver optimization possible. Powered by deep learning, natural language processing (NLP) has achieved great success in analyzing and understanding a large amount of language texts. Inspired by research results in both areas, we aim to provide a new view on communication systems from the semantic level. Particularly, we propose a deep learning based semantic communication system, named DeepSC, for text transmission. Based  on  the  Transformer, the  DeepSC  aims  at maximizing  the  system  capacity  and  minimizing  the  semantic  errors  by recovering  the  meaning  of  sentences, rather than bit- or symbol-errors in traditional communications. Moreover, transfer learning is used to ensure the DeepSC applicable to different communication environments and to accelerate the model training process. To justify the performance of semantic communications accurately, we also initialize a new metric, named sentence similarity. Compared with the traditional communication system without considering  semantic information exchange, the proposed DeepSC is more robust to channel variation and is able to achieve better performance, especially in the low signal-to-noise (SNR) regime, as demonstrated by the extensive simulation results.
\end{abstract}


.
\IEEEpeerreviewmaketitle

%\clearpage
\section{Introduction}

\IEEEPARstart{B}ASED on Shannon and Weaver~\cite{weaver1953recent},  communication could be categorized into three levels: i) transmission of symbols; ii) semantic exchange of transmitted symbols; iii) effects of semantic information exchange. The first level of communication mainly concerns about the successful transmission of symbols from the transmitter to the  receiver, where the transmission accuracy is mainly measured at the level of bits or symbols. The second level of communication deals with the semantic information sent from the transmitter and the  meaning interpreted at the receiver, named as semantic communication.  The third level deals with the effects of communication that turn into the ability of the receiver to perform certain tasks in the way desired by the transmitter.

In the past decades, communications primarily focus on how to accurately and effectively transmit symbols (measured by bits) from the transmitter to the receiver. In such systems, bit-error rate (BER) or symbol-error rate (SER) is usually taken as the performance metrics \cite{wireless1}. With the development from the first generation (1G)  to the fifth generation (5G), the achieved transmission rate has been improved tens of thousands of times and the system capacity is gradually approaching to the Shannon limit. Recently, various new applications appear, such as autonomous transportation, consumer robotics, environmental monitoring, and tele-health \cite{atzori2010internet, jameel2019internet}. The interconnection of these applications will generate a staggering amount of data in the order of zetta-bytes. Besides, these applications need to support massive connectivity over limited spectrum resources but require  lower latency, which poses critical challenges to traditional source-channel coding.  Semantic communications can process data in the semantic domain by extracting the meanings of data and filtering out the useless, irrelevant, and unessential information, which further compresses data while reserving the meanings. Moreover, semantic communication is expected to be robust to terrible channel environments, i.e., low signal-to-noise ratio (SNR) region, which fits well the applications requiring high reliability. These factors motivate us to develop intelligent communication systems by considering the semantic meaning behind digital bits to enhance the accuracy and efficiency of communications. 

Different from the conventional communications, semantic communications aim to transmit the information relevant to the transmission goal. For example, for text transmission, the meaning is thereby essential information and the expression, i.e., is expression of word, are unnecessary. By doing so, the data traffic would be reduced significantly.  Such a system could be particularly useful when the bandwidth is limited, the SNR is low, or the BER/SER is high in typical communication systems. 

Historically, the \textcolor{black}{concept} of semantic communication was developed several decades ago. Inspired by Shannon and Weaver [1], Carnap \emph{et al.} \cite{carnap1952outline} were the first to introduce the semantic information theory (SIT) based on logical probabilities ranging over the contents. Afterwards, a generic model of semantic communication (GMSC) \textcolor{black}{was} proposed as an extension of the SIT, where the concepts of semantic noise and semantic channel were first defined \cite{bao2011towards}. As pointed out in \cite{juang2011quantification}, the analysis and design of a communication system for optimal transmission of intelligence are faced with several challenges. For instance, \textit{how to define error in the intelligence transmission?} In \cite{basu2014preserving}, a lossless semantic data compression theory by applying the GMSC \textcolor{black}{was} developed, which means that data can be compressed at semantic level so that the size of the data to be transmitted can be reduced significantly. Recently, an end-to-end (E2E) semantic communication framework  integrates the semantic inference and physical layer communication problems, where the transceiver is optimized to reach the Nash equilibrium while minimizing the average semantic errors \cite{semantic2018game}. However, the semantic error in \cite{semantic2018game} measures the meaning of each word rather than the whole sentence.  These aforementioned works provide some insights and remarks for the design of semantic communications, but many issues remain unexplored.

%\subsection{Natural Language Processing}
%这个地方需要给出四个参考文献，第一个是word2vec，第二个是神经网络，第三个是seq2seq模型，第四个是pre-trained model
%DL has shown the power in semantic information processing, especially in filed of machine translation \cite{Bahdanau2014Neural} and conversational artificial intelligence (AI) \cite{Cambria2014Jumping}. Historically, the authors in \cite{mikolov2013efficient} proposed the \textit{word2vec} model to capture the relationship between words, which enables words to be represented by dense vectors rather than one-hot sparse vector. Even if these dense word vectors can capture the relationship between words, it is failed to describe syntax information. In order to solve such problem, recurrent neural networks (RNNs) are proposed for learning the syntax information \cite{graves2013generating} and these models based on RNNs, which predict future values from past values, require a big processing power. In order to reduce these costs, models use convolutional neural networks (CNNs) that are easy to parallelize, which isn’t possible with RNN \cite{KalchbrennerGB14}. \cite{wu2016google}

%\subsection{Contributions}
Recent advancements on deep learning (DL) based natural language processing (NLP) and communication systems inspire us to investigate  semantic communication to realize the second level communications as aforementioned \cite{Bahdanau2014Neural, Cambria2014Jumping, farsad2018neural, 9018199, 8444648, qin2019deep}.  The considered semantic communication system mainly focuses on the joint semantic-channel coding and decoding, which aims to extract and encode the semantic information of sentences \textcolor{black}{rather} than simply a sequence of bits or a word.  For the semantic communication system, we face the following questions:
\begin{itemize}
\item [] \hspace{-4mm}\textit{Question 1: How to define the meaning behind the bits? }
\item [] \hspace{-4mm}\textit{Question 2: How to measure the semantic error of sentences?}
\item [] \hspace{-4mm}\textit{Question 3: How to jointly design the semantic  and channel coding?}
\end{itemize} 

In this paper, we investigate the semantic communication system by applying machine translation techniques in NLP to physical layer communications. Specifically, we propose a DL enabled semantic communication system (DeepSC) to address the aforementioned challenges. The main contributions of this paper are summarized as follows:
\begin{itemize}
% 主要的目的是使得通信系统可以工作在语义角度，提供更高的鲁棒性和表现
\item Based on the Transformer~\cite{vaswani2017attention}, a novel framework for the DeepSC is proposed, which can effectively extract the semantic information from texts with robustness to noise. In the proposed DeepSC, a joint semantic-channel coding is designed to cope with channel noise and semantic distortion, which addresses aforementioned \textit{Question 3}.

%the transmitter can extract semantic information from text and the receiver can 

\item The transceiver of the DeepSC is composed of  semantic encoder, channel encoder, channel decoder, and semantic decoder. To understand the semantic meaning as well as maximize the system capacity at the same time,  the receiver is optimized with two loss functions:  cross-entropy and mutual information. Moreover, a new metric is proposed to accurately reflect the performance of the DeepSC at the semantic level. These  address the aforementioned \textit{Questions 1} and \textit{2}.

\item To make the DeepSC applicable to various communication scenarios, deep transfer learning is adopted to accelerate the model re-training. With the re-trained model, the DeepSC can recognise various knowledge input and recover semantic information from distortion.

\item Based on extensive simulation results, the proposed DeepSC outperforms the traditional communication system and improves the system robustness at the low SNR regime.

\end{itemize}

The rest of this paper is organized as follows. Related work is briefly reviewed in Section II. The framework of a semantic communication system is presented and a corresponding problem is formulated  in Section III. Section IV details the proposed DeepSC and extends it to dynamic environments. Numerical results are presented  in Section VI to show the performance of the DeepSC. Finally, Section VII concludes this paper.

$Notation$: $\mathbb{C}^{n \times m}$ and $\mathbb{R}^{n \times m}$ represent \textcolor{black}{sets} of complex and real matrices of size  $n \times m$, respectively. Bold-font variables denote matrices or vectors. $x \sim {\cal CN}(\mu,\sigma^2)$ means variable $x$ follows a circularly-symmetric complex Gaussian distribution with mean $\mu$ and covariance $\sigma^2$. $(\cdot)^T$ and $(\cdot)^H$ denote the transpose and Hermitian, respectively. $\Re \{\cdot \}$ and $\Im \{\cdot \}$ refer to the real and imaginary parts of a complex number.  Finally, ${\bf a}\otimes{\bf b}$ indicates the inner product of vectors $\bf a$ and $\bf b$.

\section{Related Work}
This section provides a brief review of the related work on the E2E physical layer communication systems and the deep neural network (DNN)  techniques adopted in NLP.
%直接加NLP的东西
%\textcolor{black}{As shown in Fig. 2,} ${\cal K}_t$ and ${\cal K}_r$ in transmitter and receiver consist of different sentences respectively, which provide the additional semantic information to facilitate the transmissions. Meanwhile, semantic coding process can  losslessly compress information by language model while the meaning of semantic information does not change. 
%要提到语义，语法，句法，然后给出深度神经网络解决这些问题的方法，RNN, CNN, Transformer
\subsection{End-to-End Physical Layer Communication Systems}

DL techniques have shown great potential in processing various intelligent tasks, i.e., computer vision and NLP. Meanwhile, it is possible to train neural networks and run them on mobile devices due to the increasing hardware computing capability. In the communication area, some pioneering works have been carried on DL based E2E physical layer communication systems, which merge the blocks in \textcolor{black}{traditional} communication systems \cite{o2017introduction, DornerCHB18, aoudia2019model, Hye2020, park2020end,  gold2018, bourtsoulatze2019deep}. By adopting the structure of autoencoder in DL and removing block structure, the transmitter and receiver in the E2E system are optimized jointly as an E2E reconstruction task. It has been demonstrated that such an E2E system outperforms uncoded binary phase shift keying (BPSK) and Hamming coded BPSK in terms of BER \cite{o2017introduction}. Besides, there are several initial works on dealing with the missing channel gradient during training.  A DNN based two-phase of training processing has been proposed, where the transceiver is trained by an stochastic channel model and the receiver is fine-tuned under real channels \cite{DornerCHB18}. Reinforcement learning has been exploited in \cite{aoudia2019model} to acquire the channel gradient under an unknown channel model, which achieves better performance than the differential quadrature phase-shift keying (DQPSK) over real channels. A conditional generative adversarial net (GAN) has been applied in \cite{Hye2020} to use a DNN to represent the channel distortion so that the gradients can  pass through a unknown channel to the transmitter DNN during the training of the E2E communication system. Meta-learning combined with a limited number of pilots has been developed for training the transceiver and enables the fast training of network with less amount of data \cite{park2020end}. 

%Taking variable input length into consideration, an convolutional autoencoder has been proposed for the joint design of transmitter and receiver, which achieves better performance than \textcolor{black}{the} traditional methods over fading channels with low complexity \cite{zhu2019joint}. 

Considering the types of sources, the joint source-channel coding for texts \cite{gold2018}  and images~\cite{bourtsoulatze2019deep} aims to recover the source information at the receiver directly rather than the digital bits. Meanwhile, traditional metrics, such as BER, cannot reflect the performance for such systems well. Therefore, word-error rate and peak signal-to-noise ratio (PSNR) are  adopted for measuring the accuracy of source information recovery.

\subsection{Semantic Representation in Natural Language Processing}
%加些词向量的工作进来
NLP makes machines understand human languages, with the main goal to understand the syntax and text. Initially, natural language can be described by the joint probability model according to the context \cite{KneserN95}. Thus, language models provide context to distinguish words and phrases that have similar semantic meaning. Although such NLP technologies based on statistical model are developed to describe the \textcolor{black}{probability of a certain word coming after another in a sentence}, it is hard to deal with long sentences, i.e. the ones over 15 words, and the syntax. To understand long sentences, the \textit{word2vec} model in \cite{mikolov2013efficient} captures the relationship among words, which makes similar words ending up with a closer distance in the vector space. Even if these dense word vectors can capture the relationship among words, they fail to describe syntax information. In order to solve such problems, the underlying meaning of texts is represented by using various DL techniques, which is able to extract the semantic information in long sentences and their syntax. A deep contextualized word representation has been proposed in \cite{peters2018deep}, which models both complex characteristics of word usages, e.g., syntax and semantics, and how these usages vary across linguistic contexts (i.e., to model polysemy). However, the above word representation approaches are designed for specific tasks and may need to be redesigned whenever the task changes. In \cite{DevlinCLT19}, a general word representation model, named bidirectional encoder representations from transformers (BERT), has been developed to provide word vectors for various NLP tasks without requiring redesign of word representations.

\subsection{Comparison of State-of-Art NLP Techniques}
There are three types of neural networks used for NLP tasks, including recurrent neural networks (RNNs), convolutional neural networks (CNNs) and  fully-connected neural networks (FCNs)  \cite{schuster1997bidirectional}. By introducing  RNNs, language models can learn the whole sentences and capture the syntax information effectively \cite{graves2013generating}. However, for long sentences, particularly, the distance between subject and predicate is more than 10 words, RNNs cannot find the correct subject and predicate. For example, for sentence “the person who works in the new post office is walking to the store”, RNNs fail to recognise the relationship between “the person” and “is”. Besides, because of linear sequence structure, RNNs lack of parallel computing capability, which means that RNNs are time-consuming. CNNs were born with the capability of parallel computing \cite{KalchbrennerGB14}. However, even if CNNs can use deeper network to extract semantic information in long sentences, its performance is not as good as that of RNNs because the kernel size in CNNs is small to guarantee the computational efficiency. By combining with the attention mechanism,  language models based on FCNs, such as Transformer \cite{vaswani2017attention}, pay more attention to the useful semantic information for performance improvement on various NLP tasks.  It is worth noting that the Transformer has the advantages of both RNNs and CNNs \cite{vaswani2017attention}. Particularly,  the self-attention mechanism is adopted, which enables the models to understand sentences regardless of \textcolor{black}{their} lengths.

\section{System Model and Problem Formulation}
The considered system model consists of two levels: semantic level and transmission level, as shown in Fig. \ref{system model}. The semantic level addresses semantic information processing for encoding and decoding to extract the semantic information. The transmission level guarantees that semantic information can be exchanged correctly over the transmission medium. Overall, we consider an intelligent E2E communication system with the stochastic physical channel, where the transmitter and the receiver have certain background knowledge, i.e., different training data. The background knowledge could be various for different application scenarios. 
\begin{defi}
	Semantic noise is a type of disturbance in the exchange of a message that interferes with the interpretation of the message due to ambiguity in words, a sentence or symbols used in the message transmission.
\end{defi}

\begin{defi}
	Physical channel noise is caused by the physical channel impairment, such as, additive white Gaussian noise (AWGN), fading channel, and multiple path, which incurs the signal attenuation and distortion.
\end{defi}

\subsection{Problem Description}
\begin{figure}
	\centering
	\includegraphics[width=90mm]{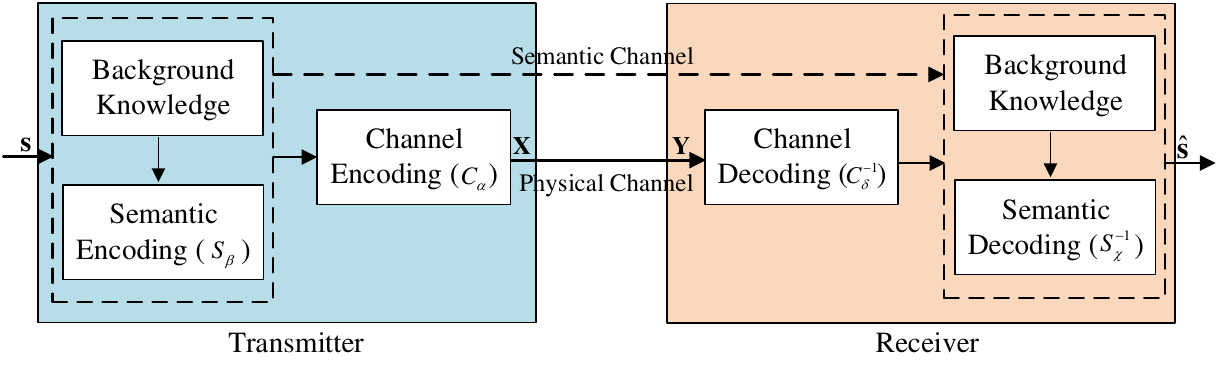}
	\caption{The framework of proposed DL enabled semantic communication system, DeepSC.}
	\label{system model}
\end{figure}

As in Fig. \ref{system model}, the transmitter maps a sentence, $\bf s$, into a complex symbol stream, $\bf x$, and then passes it through the physical channel with transmission impairments, such as distortion and noise. The received, $\bf y$,  is decoded at the receiver to estimate the original sentence, ${\bf s}$.  We jointly design the transmitter and receiver with DNNs since DL enables us to train a model with inputting variable-length sentences and different languages.

Particularly, we assume that the input of the DeepSC is a sentence, ${\mathbf{s}} = \left[ {{w_1},{w_2}, \cdots ,{w_L}} \right]$, where $w_l$ represents the $l$-th word in the sentence. As shown in Fig. \ref{system model}, the transmitter consists of two parts, named semantic encoder and channel encoder, \textcolor{black}{to} extract the semantic information from $\bf s$ and guarantee successful transmission of semantic information  over the physical channel. The encoded symbol stream can be represented by
\begin{equation}
	{\mathbf{x}} = {C_{{\bm{ {\bm \alpha} }}}}\left( {{S_{{{\bm \beta} }}}\left( {\mathbf{s}} \right)} \right),
\end{equation}
where ${\bf{x}} \in {\mathbb{C}^{M \times 1}}$, ${S_{ {\bm \beta}} }\left( \cdot \right)$ is the semantic encoder network with the parameter set ${{\bm \beta}}$ and ${C_{\bm \alpha} }\left( \cdot \right)$  is the channel encoder with the parameter set $\mathbf{ {\bm \alpha}}$. In order to simplify the analysis, we assume the coherent time is $M$. If $\bf x$ is sent,  the signal received at the receiver will be
\begin{equation}
	{\bf{y}} = h{\bf{x}} + {\bf{n}},
\end{equation}
where ${\bf{y}} \in {C^{M \times 1}}$, $h$ represents the Rayleigh fading channel with ${\cal CN}\left( {0,1} \right)$ and ${\bf n} \sim {\cal CN}\left( {0,\sigma _n^2} \right)$. For E2E training of the encoder and the decoder, the channel must allow back-propagation. Physical channels can be formulated by neural networks. For example, simple neural networks could be used to model the AWGN channel, multiplicative Gaussian noise channel, and the erasure channel \cite{gold2018}. While for the fading channels, more complicated neural networks are required \cite{Hye2020}. In this paper, we mainly consider the AWGN channels and Rayleigh fading channels for simplicity while focus on semantic coding and decoding.

As shown in Fig. \ref{system model}, the receiver includes channel decoder and semantic decoder to recover the transmitted symbols and then transmitted sentences, respectively. The decoded signal can be represented as 
\begin{equation}
	{\mathbf{\hat s}} = {S^{-1}_{\bm \chi} }\left( {{C^{-1}_{\bm \delta} }\left( {\mathbf{y}} \right)} \right),
\end{equation}
where the $\bf \hat s$ is the recovered sentence, $C^{-1}_{\bm \delta} \left( \cdot \right)$ is the channel decoder with the parameter set ${\bm \delta}$ and $S^{-1}_{\bm \chi} \left( \cdot \right)$ is the semantic decoder network with the parameter set $\mathbf{ {\bm \chi}}$. 

The goal of the system is to minimize the semantic errors while reducing the number of symbols to be transmitted. However, we face two challenges in the considered system. The first challenge is how to design joint semantic-channel coding. The other one is semantic transmission, which has not been considered in the traditional communication system. Even if the existing communication system can achieve a low BER, several bits, distorted by the noise and beyond error correction capability, could lead to understanding difficulty as the partial semantic information of the whole sentence might be missed. In order to achieve successful recovery at semantic level, we design  semantic and channel coding jointly in order to keep the meaning between $\hat {\bf s}$ and $\bf s$ unchanged, which is enabled by a new DNN framework. The cross-entropy (CE) is used as the loss function  to measure the difference between $\bf s$ and $\bf \hat  s$, which can be formulated as
\begin{equation}\label{loss function 1}
    \begin{aligned}
     &{\cal L}_{\rm CE}({\mathbf{s}},{\mathbf{\hat s}}; {\bm \alpha}, {\bm \beta}, {\bm \chi}, {\bm \delta}) =  \\
     &- \sum\limits_{l = 1} {{q\left( {{w_l}} \right)}\log \left( {p\left( {{w_l}} \right)} \right)}  + (1 - {q\left( {{w_l}} \right)})\log \left( {1 - p\left( {{w_l}} \right)} \right),
     \end{aligned}
\end{equation}
where $q(w_l)$ is the real probability that the  $l$-th word, $w_l$, appears in  estimated sentence $\bf s$, and  $p({w_l})$ is the predicted probability that the  $i$-th word, $ w_i$, appears in  sentence $\hat {\bf s}$. The CE can measure the difference between two probability distributions. Through reducing the loss value of CE, the network can learn the word distribution, $q(w_l)$, in the source sentence, $\bf s$, which indicates that the syntax, phrase, the meaning of words in context can be learnt by the network.
Besides, jointly designing and training semantic-channel coding can make the whole network learning the knowledge for the specific goal. In other words, the channel coding can pay more attention in protecting the semantic information related to transmission goal while neglecting other irrelevant information. Separately designing will make channel coding addressing all information equally.

\subsection{Channel Encoder and Decoder Design}
One important goal \textcolor{black}{on} designing a communication system is to maximize the capacity \textcolor{black}{or} the data transmission rate. Compared with BER, the mutual information can provide extra information to train a receiver. The mutual information of the transmitted symbols, $\bf x$, and the received symbols, $\bf y$,  can be computed by
\begin{equation}\label{mutual information 1}
\begin{aligned}
  I\left( {\bf x;y} \right) &= \int {_{ {\mathcal {X}}\times {\mathcal {Y}}}p\left( {x,y} \right)\log \frac{{p\left( {x,y} \right)}}{{p\left( x \right)p\left( y \right)}}} dxdy \\
  &= {\mathbb{E}}_{p(x,y)}\left[ {\log \frac{{p\left( {x, y} \right)}}{{p\left( y \right)p\left( x \right)}}} \right],
 \end{aligned}
\end{equation}
where  $(\bf x, \bf y)$ is a pair of random variables with values over the space ${ {\mathcal {X}}\times {\mathcal {Y}}}$, where $\cal X$ and $\cal Y$ are the  spaces for $\bf x$ and $\bf y$.  $p(x)$ and $p(y)$ are the marginal probability of sending $\bf x$ and received $\bf y$, respectively, and $p(x,y)$ is the joint probability of $\bf x$ and $\bf y$. The mutual information is equivalent to the Kullback-Leibler (KL) divergence between the marginal probabilities and the joint probability, which is given by

\begin{equation}\label{6}
   I\left( {\bf x;y} \right) = {D_{\rm KL}}\left( {p\left( {x,y} \right)\left\| {p\left( x \right)p\left( y \right)} \right.} \right).
\end{equation}
\textcolor{black}{From  \cite{belghazi2018mine}, we have the following theorem,}

\begin{theorem}\label{theorem}
The KL divergence admits the following dual representation
\begin{equation}\label{14}
    {D_{\rm KL}}\left( {P\left\| Q \right.} \right) = \mathop {\sup }\limits_{T:\Omega  \to R} {E_P}\left[ T \right] - \log \left( {{E_Q}\left[ {{e^T}} \right]} \right),
\end{equation}
where the supremum is taken over all functions $T$ such that the two expectations are finite.
\end{theorem}

According to Theorem \ref{theorem}, the KL divergence can also be represented as
\begin{equation}\label{15}
  {D_{\rm KL}}\left( {p\left( {x,y} \right)\left\| {p\left( x \right)p\left( y \right)} \right.} \right) \geqslant {{\mathbb E}_{p\left( {x,y} \right)}}\left[ T \right] - \log \left( {{{\mathbb E}_{p\left( x \right)p\left( y \right)}}\left[ {{e^T}} \right]} \right).
\end{equation}
Thus, the lower bound of $I\left( {\bf x;y} \right)$ can be obtained from \eqref{6} and \eqref{15}. In order to find a tight bound on the $I\left( {\bf x;y} \right)$, an unsupervised method is used to train the network $T$, where $T$ can be approximated by neural network. Meanwhile, the expectation in \eqref{15} can be computed by sampling, which converges to the true value as the number of samples increases. Then, we can optimize the encoder by maximizing the mutual information defined in \eqref{15} and the related loss function can be given by 
\begin{equation}\label{eq14}
{\cal L}_{\rm MI}({\bf x,y};T) = {{\mathbb E}_{p\left( {x,y} \right)}}\left[ f_T \right] - \log \left( {{{\mathbb E}_{p\left( x \right)p\left( y \right)}}\left[ {{e^{f_T}}} \right]} \right),
\end{equation}
where $f_T$ is composed by a neural network, in which the inputs are samples from $p(x,y)$, $p(x)$, and $p(y)$. In our proposed design,  $\bf x$ is generated by the function $C_{\bm \alpha}$ and $S_{\bm \beta}$, thus the loss function can be represented by ${\cal L}_{\rm MI}({\bf x,y};T, {\bm \alpha}, {\bm \beta})$ with
\begin{equation}\label{mutual information}
     {\cal L}_{\rm MI}({\bf x,y};T, {\bm \alpha}, {\bm \beta}) \leqslant  I(\bf x;y) .
\end{equation}

From \eqref{mutual information}, the loss function can be used to train neural networks to get ${\bm \alpha}$, ${\bm \beta}$, and $T$. For example, the mutual information can be estimated by training network $T$ when the encoders $\bf {\bm \alpha}$ and $\bf {\bm \beta}$ are fixed. Similarly, the encoder can be optimized by training ${\bm \alpha}$ and ${\bm \beta}$ when the mutual information is obtained.

% \subsubsection{Document similarity}
\subsection{Performance Metrics}
Performance criteria are important to the system design. In the E2E communication system, the BER is usually taken as the training target by the transmitter and receiver, which sometimes neglects the other aspect goals of communication. For text transmission, BER cannot reflect performance well. Except from human judgement to establish the similarity between sentences, bilingual evaluation understudy (BLEU) score is usually used to measure the results in machine translation \cite{papineni2002bleu}, which will be used as one of the performance metrics in this paper. However, the BLEU score can only compare the difference between words in two sentences rather than their semantic information. Therefore, we initialize a new metric, named sentence similarity, to describe the similarity level of two sentences in terms of their semantic information, which is introduced in the following. This provides a solution to \textit{Question 2}.

\subsubsection{BLEU Score}
Through counting the difference of $n$-grams between transmitted and received texts, where $n$-grams means that the size of a word group. For example, for sentence ``weather is good today", $1$-gram: ``weather", ``is", ``good" and ``today", $2$-grams:  ``weather is", ``is good" and ``good today". The same rule applies for the rest. 

For the transmitted sentence $\bf s$ with length $l_{\bf s}$ and the decoded sentence $\bf \hat s$ with length $l_ {\bf \hat s}$, the BLEU can be expressed as
\begin{equation}
  \log {\text{BLEU = min}}\left( {1 - \frac{l_{\bf \hat s}}{l_{\bf s}},0} \right) + \sum\limits_{n = 1}^N {{u_n}\log {p_n}}, 
\end{equation}
where $u_n$ is the weights of $n$-grams and $p_n$ is the $n$-grams score, which is
\begin{equation}\label{wer}
{p_n} = \frac{{\sum\nolimits_k {\min \left( {{C_k}\left( {{\mathbf{\hat s}}} \right),{C_k}\left( {\mathbf{s}} \right)} \right)} }}{{\sum\nolimits_k {\min \left( {{C_k}\left( {{\mathbf{\hat s}}} \right)} \right)} }},
\end{equation}
where $C_k(\cdot)$ is the frequency count function for the $k$-th elements in $n$-th grams.

The output of BLEU is a number between 0 and 1, which indicates how similar the decoded text is to the transmitted text, with 1 representing highest similarity. However, few human translations will attain the score of 1 since word error may not make the meaning of a sentence different. For instance, the two sentences, ``my car was parked there" and ``my automobile was parked there",  have the same meaning but with different BLEU scores since they use different words. To characterize such a feature, we propose a new metric, the sentence similarity, at the sentence level in addition to the BLEU score.
\begin{figure*}[t!]
	\centering
	\includegraphics[width=165mm, height = 35mm]{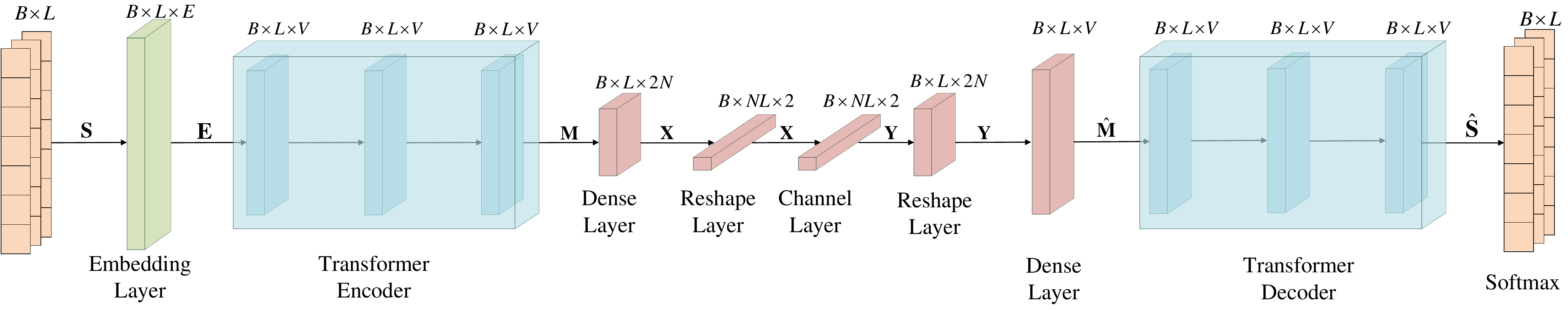}
	\caption{The proposed neural network structure for the semantic communication system.}
	\label{network structure}
\end{figure*}
\subsubsection{Sentence Similarity}
\textcolor{black}{A word} can take different meanings in different contexts. For instance, the meanings of mouse in biology and machine \textcolor{black}{are} different. The traditional method, such as \textit{word2vec} \cite{mikolov2013efficient}, cannot recognise the polysemy, of which the problem is how to use an  numerical vector to express the word while the numerical vector varies in different contexts. According to the semantic similarity, we propose to calculate the sentence \textcolor{black}{similarity between} the original sentence, $\bf s$, and the recovered sentence, $\bf \hat s$, as
\begin{equation}\label{ss}
   {\text{match}}\left( {\bf \hat s,s} \right) =  \frac{{{\bm B_{\bm \Phi} }\left( {\bf{s}} \right) \cdot {\bm B_{\bm \Phi} }{{\left( {{\bf{\hat s}}} \right)}^T}}}{{\left\| {{\bm B_{\bm \Phi} }\left( {\bf{s}} \right)} \right\|\left\| {\bm {B_{\bm \Phi} }\left( {{\bf{\hat s}}} \right)} \right\|}},
\end{equation}
%\textcolor{black}{the score between 0 and 1,}
where $ {\bm B_{\bm \Phi} }$, representing BERT  \cite{DevlinCLT19},  is a huge pre-trained model including billions of parameters used for extracting the semantic information.  The sentence similarity defined in \eqref{ss} is  a number between 0 and 1, which indicates how similar the decoded sentence  is  to  the  transmitted sentence,  with  1  representing  highest  similarity and 0 representing no similarity between $\bf s$ and $\bf \hat s$.

Compared with BLEU score, BERT has been fed by billions of sentences. Therefore, it has already learnt the semantic information from these sentences and can generate different semantic vectors in different contexts effectively.  With the BERT, the semantic information behind a transmitted sentence, $\bf s$, can be expressed as $\bf c$. Meanwhile, the semantic information conveyed by the estimated sentence is expressed as $\bf \hat c$. For $\bf c$ and $\bf \hat c$, we can compute the sentence similarity by $\text{match}(\bf c,\bf \hat c)$. 

\section{Proposed Deep Semantic Communication Systems}
%.  The network structure shows in Fig. \ref{network structure},
In this section, we propose a DNN for the considered semantic communication system, named as DeepSC, of which the Transformer is adopted for text understanding. Then, transfer learning is adopted to make the DeepSC applicable \textcolor{black}{to} different background knowledge and dynamic communication environments. This provides the solutions to \textit{Question 1,3}.

\subsection{Basic Model}
The proposed  DeepSC is as shown in Fig \ref{network structure}. Particularly, the transmitter consists of a semantic encoder to extract \textcolor{black}{the} semantic features from the texts to be transmitted and a channel encoder to generate symbols to facilitate the transmission \textcolor{black}{subsequently}. The semantic encoder includes multiple Transformer encoder layers and the channel encoder uses dense layers with different units. The AWGN channel is interpreted as one layer in the model. Accordingly, the DeepSC receiver is composited with a channel decoder for symbol detection and a semantic decoder for text estimation, the channel decoder includes dense layers with different units and the semantic decoder includes multiple Transformer decoder layers.  The loss function can be expressed as
\begin{equation}\label{total loss}
   {\cal L}_{\rm total} = {\cal L}_{\rm CE}({\mathbf{s}},{\mathbf{\hat s}}; {\bm \alpha}, {\bm \beta}, {\bm \chi}, {\bm \delta}) - \lambda {\cal L}_{\rm MI}({\bf x,y};T, {\bm \alpha}, {\bm \beta}),
\end{equation}
where the first term is the loss function considering the sentence similarity, which aims to minimize the semantic difference between $\bf s$ and $\bf \hat s$ by training the whole system. The second one is the loss function for  mutual information, which maximize the achieved data rate during the transmitter training. The parameter $\lambda$ ($0 \le \lambda \le 1$) is the weight for the second term.

\begin{figure}
	\centering
	\includegraphics[width=1.6in]{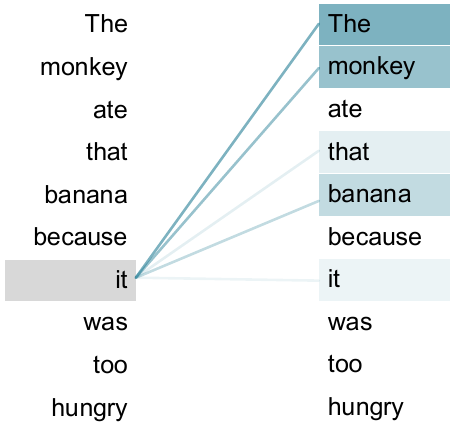}
	\caption{An example of the self-attention mechanism following long-distance dependency in the Transformer encoder.} 
	\label{attention}
\end{figure}

The core of Transformer is the multi-head self-attention mechanism, which enables the Transformer to view the previous  predicted word in the sequence, thereby better predicting the next word. Fig. \ref{attention} gives an example of the self-attention mechanism for the word ‘it’. From Fig.~\ref{attention}, attention attend to a distant dependency of the pronoun, ‘it’, completing pronoun reference ``the animal'', which demonstrates that the self-attention mechanism can learn the semantic and therefore solve  aforementioned \textit{Question 1}. 

\begin{algorithm}
	\caption{DeepSC network training algorithm.} 
	\label{alg:Framwork} 
	\text{\textbf{Initialization}: Initial the weights $\bf W$ and bias $\bf b$.}
	
	\begin{algorithmic}[1] 
	\STATE \text{\textbf{Input}: The background knowledge set $\cal K$.  }
	\STATE Create the index to words and words to index, and then embedding words.
	\WHILE{Stop criterion is not met}
	\STATE Train the mutual information estimated model.
	\STATE Train the whole network.
	\ENDWHILE
	\STATE \text{\textbf{Output}: The whole network $S_{\bm \beta} (\cdot), C_{\bm \alpha} (\cdot), C^{-1}_{\bm \delta} (\cdot), S^{-1}_{\bm \chi} (\cdot)$. }
	\end{algorithmic}
\end{algorithm}

As shown in Algorithm \ref{alg:Framwork}, the training process of the DeepSC consists of two phases due to different loss functions. After initializing the weights, $\bf W$, bias, $\bf b$, and using embedding vector to represent the input words, the first phase is to train the mutual information model by unsupervised learning to estimate the achieved data rate for the second phase. The second phase is to train the whole system with \eqref{total loss} as the loss function.  Each phase aims to minimize the loss by gradient descent with mini-batch until the stop criterion is met, the max number of iteration is reached, or none of terms in the loss function is  decreased any more. Different from performing semantic coding and channel coding separately, where the channel encoder/decoder will deal with the digital bits rather than the semantic information, the joint semantic-channel coding can  preserve semantic information when compressing data, which provides the detailed solution for  aforementioned \textit{Question 3}. The two training phases are described in the following:

\begin{figure}
	\centering
	\includegraphics[width=75mm]{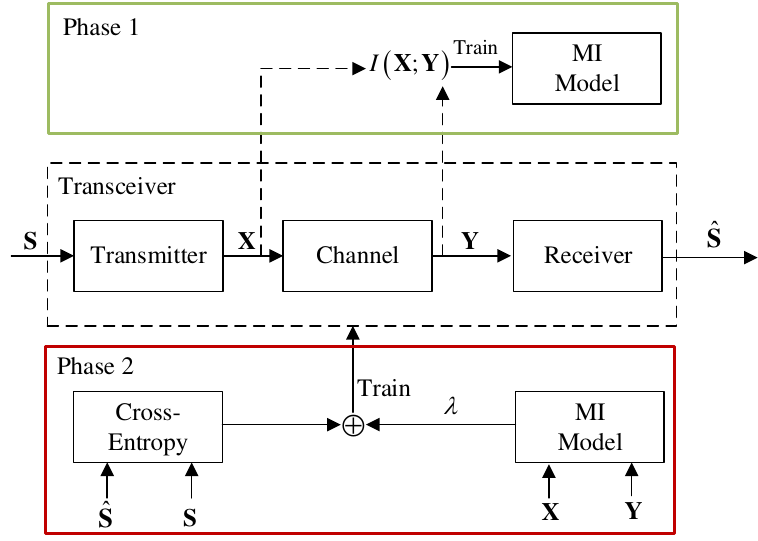}
	\caption{The training framework of the DeepSC: phase 1 trains the mutual information estimation model; phase 2 trains the whole network based on the cross-entropy and mutual information.}
	\label{Training framework}
\end{figure}

\subsubsection{Training of mutual information estimation model}
The mutual information estimation model training process is illustrated in Fig. \ref{Training framework} and the pseudocode is given in Algorithm \ref{channel encoder function}. First, the knowledge set ${\cal K}$ generates a minibatch of sentences ${\mathbf{S}} \in {\Re ^{B \times L \times 1}}$, where $B$ is the batch size, $L$ is the length of sentences. Through the embedding layer, the sentences can be represented as a dense word vector ${\bf E}  \in {\Re^{B \times L \times E}}$, where $E$ is the dimension of the word vector. Then, pass the semantic encoder layer \textcolor{black}{to obtain ${\bf M} \in {\Re^{B \times L \times V}}$}, the semantic information conveyed by $\bf S$, where $V$ is the dimension of Transformer encoder's output. Then, $\bf M$ is encoded into symbols ${\bf X}$ to cope with the effects from the physical channel, where $ {\bf X} \in {\Re ^{B \times NL \times 2}}$. After passing through the channel, the receiver obtains signal $\bf Y$ distorted by the channel noise. Based on \eqref{eq14}, the loss, ${\cal L}_{\rm MI}({\bf X},{\bf Y};T,{\bm \alpha}, {\bm \beta})$, can be computed based on the transmitted symbols, $\bf X$, and the received symbols, $\bf Y$, under the AWGN channels. Finally, according to  computed ${\cal L}_{\rm MI}$, the stochastic gradient descent (SGD) is exploited to optimize the weights and bias of $f_T(\cdot)$.
% is the matrix containing $LN$ symbols and each word is represented by $N$ symbols

\begin{algorithm}[htbp]
	\caption{Train mutual information estimation model.} 
	\label{channel encoder function} 
	\begin{algorithmic}[1] %这个1 表示每一行都显示数字
	\STATE \text{\textbf{Input}: The knowledge set $\cal K$.}
	\STATE \text{\textbf{Transmitter}}:
	\STATE \qquad BatchSource(${\cal K}$) $\to$ $\bf S$.
	\STATE  \qquad $S_{\bm \beta} (\bf S)\to {\bf M}$.
	\STATE \qquad $C_{\bm \alpha}(\bf M) \to {\bf X}$.
	\STATE \qquad Transmit $\bf X$ over the channel.
	\STATE \text{\textbf{Receiver}}:
	\STATE \qquad Receive $\bf Y$.
	\STATE \qquad Compute loss ${\cal L}_{\rm MI}$ by \eqref{eq14}.
	\STATE \qquad Train $T$ $\to$ Gradient descent ($T, {\cal L}_{\rm MI}$).
	\STATE \text{\textbf{Output}: The mutual information estimated model $f_T (\cdot)$.}
	\end{algorithmic}
\end{algorithm}

\subsubsection{Whole network training}
The whole network training process is illustrated in Algorithm \ref{Train Whole Network Function}. First, minibatch $\bf S$ from  knowledge $\cal K$ is encoded into $\bf M$ at the semantic level, then $\bf M$ is encoded into symbol ${\bf X}$ for transmission over \textcolor{black}{the physical} channels. At the receiver, \textcolor{black}{distorted} symbols $\bf Y$ are received \textcolor{black}{and}  then decoded by the channel decoder layer, where ${\hat {\bf M}} \in {\Re^{B \times L \times V}}$ is the recovered semantic information of the sources. Afterwards, the transmitted sentences are estimated by the semantic decoder layer. Finally, the whole network is optimized by the SGD, where the loss is computed by \eqref{total loss}.

\begin{algorithm}
	\caption{Train the whole network.} 
	\label{Train Whole Network Function} 
	\begin{algorithmic}[1] %这个1 表示每一行都显示数字
	\STATE \text{\textbf{Input}: The knowledge set $\cal K$}.
	\STATE \text{\textbf{Transmitter}}:
	\STATE \qquad BatchSource(${\cal K}$) $\to$ $\bf S$.
	\STATE  \qquad $S_{\bm \beta} (\bf S)\to {\bf M}$.
	\STATE \qquad $C_{\bm \alpha}(\bf M) \to {\bf X}$.
	\STATE \qquad Transmit $\bf X$ over the channel.
	\STATE \text{\textbf{Receiver}}:
	\STATE \qquad Receive $\bf Y$.
	\STATE \qquad $C^{-1}_{\bm \delta} (\bf Y)\to \hat{\bf M}$.
	\STATE \qquad $S^{-1}_{\bm \chi} ({\hat {\bf M}})\to \hat {\bf S}$.
	\STATE \qquad Compute loss function ${\cal L}_{\rm total} $ by \eqref{total loss}.
	\STATE \qquad Train ${\bm \beta}, {\bm \alpha}, {\bm \delta}, {\bm \chi}$  $\to$ Gradient descent (${\bm \beta}, {\bm \alpha}, {\bm \delta},$ $ {\bm \chi}, {\cal L}_{\rm total}$).
	\STATE \text{\textbf{Output}: The whole network $S_{\bm \beta} (\cdot), C_{\bm \alpha} (\cdot), C^{-1}_{\bm \delta} (\cdot), S^{-1}_{\bm \chi} (\cdot)$.}
	\end{algorithmic}
\end{algorithm}

\subsection{Transfer Learning for Dynamic Environment}
%channel 或者 knowledge
%\begin{equation}
%\begin{aligned}
%  {{\cal K}_t} \cup {{\cal K}_r} &= {\cal K} \hfill \\
%  {{\cal K}_t} \cap {{\cal K}_r} &\ne \emptyset  \hfill \\ 
%\end{aligned} 
%\end{equation}
%where ${\cal K}_t$ and ${\cal K}_r$ just share part of knowledge. 
\begin{figure}[t!]
	\centering
	\includegraphics[width=80mm]{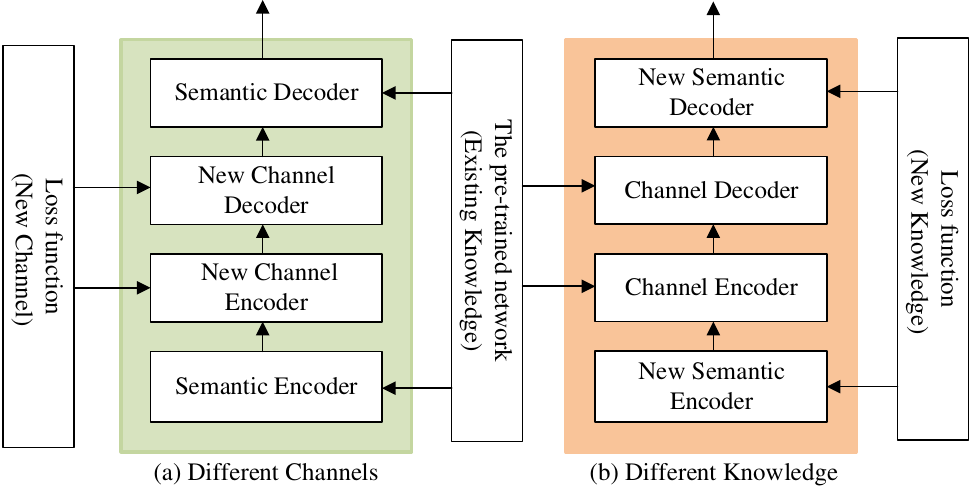}
	\caption{Transfer learning based training framework: (a) re-train channel encoder and decoder for different channels; (b) re-train semantic encoder and decoder for different background knowledge. }
	\label{Difference framework}
\end{figure}

%First, the knowledge ${\cal K}_t$ is fed to train the network 1 by channel encoder function and whole network function. Second, using the knowledge ${\cal K}_r$ trains network 2 under mutual information model function and whole network function. Then, we totally have two different network, where the network 1 includes $f_{\bm \beta}^{(t)}, f_{\bm \alpha}^{(t)}, f_{\bm \delta}^{(t)}, f_{\bm \chi}^{(t)}$ and the network 2 includes $f_{\bm \beta}^{(r)}, f_{\bm \alpha}^{(r)}, f_{\bm \delta}^{(r)}, f_{\bm \chi}^{(r)}$. Finally, the final network consists of the receiver $f_{\bm \beta}^{(t)}, f_{\bm \alpha}^{(t)}$ of network 1 and the transmitter $f_{\bm \delta}^{(r)}, f_{\bm \chi}^{(r)}$ of network 2.

\begin{algorithm}[htbp] 
	\caption{Transfer learning based training for dynamic environment.}
	\label{difference} 
%	\text{\qquad \, \,} \\
	\text{\textbf{Initialization}: Load the pre-trained model $S_{\bm \beta} (\cdot), C_{\bm \alpha} (\cdot)$,}\\
	\text{$C^{-1}_{\bm \delta} (\cdot), S^{-1}_{\bm \chi} (\cdot)$}.
	
	\begin{algorithmic}[1] %这个1 表示每一行都显示数字
		\ENSURE Training for different background knowledge\\ %算法的输出：Output
		\STATE \text{\textbf{Input}: The different background knowledge set ${\cal K}_1$ }.
		\STATE Freeze $C_{\bm \alpha} (\cdot)$ and $C^{-1}_{\bm \delta} (\cdot)$.
		\STATE Redesign  and train part of $S_{\bm \beta} (\cdot)$ and $S^{-1}_{\bm \chi} (\cdot)$.
	    \WHILE{Stop criterion is not met}
	    \STATE Train the mutual information estimated model.
	    \STATE Train the whole network.
	    \ENDWHILE\\
	    \STATE \text{\textbf{Output}: The adopted whole network.} 
%	 \end{algorithmic}
	 \vspace{0.55em}
%	 \begin{algorithmic}[1]
	    \ENSURE Training for different channel conditions\\ %算法的输出：Output
	    \STATE \textbf{Input}: The background knowledge set $\cal K$ with the different channel parameters.
	    \STATE Freeze $S_{\bm \beta} (\cdot)$ and $S^{-1}_{\bm \chi} (\cdot)$.
	    \STATE Redesign and re-train part of $C_{\bm \alpha} (\cdot) $ and $ C^{-1}_{\bm \delta} (\cdot)$. 
	    \WHILE{Stop criterion is not met}
	    \STATE Train the mutual information estimated model.
	    \STATE Train the whole network.
	    \ENDWHILE
	    \STATE \text{\textbf{Output}: The re-trained network.  } 
	\end{algorithmic}
	
\end{algorithm}
% 这一节要不要做迁移学习是个问题，需要再仔细思考一下。
% 经过仿真分析，different knowledge是不可行的，准备做transfer learning
In practice, different communication scenarios result in the different channels and the training data. However, the re-training of transmitter and receiver to meet the requirements of dynamic scenarios introduces extra costs. To address this, a deep transfer learning approach is adopted, which focuses on storing knowledge gained while solving a problem and applying it to a different but related problem. 

%the structure of network is the same as the Table \ref{table 2}

%Under difference knowledge, . However, the difference is that the transmitter and the receiver will be trained by the knowledge ${\cal K}_t$ and knowledge ${\cal K}_r$, respectively. Thus, the weights and bias in receiver and transmitter are different from shared knowledge. Notice that when we start training our network, two networks including receiver and transmitter will be trained. Then, the receiver from network 1 and the transmitter from network 2 will be combined to form the new network. 

The training process of adopting transfer learning is illustrated in Fig. \ref{Difference framework} and the pseudocode is given in Algorithm \ref{difference}, where the training modules, mutual information estimation model training, and whole network training, are the same as Algorithm \ref{channel encoder function} and Algorithm \ref{Train Whole Network Function}. First, load the pre-trained transmitter and receiver based on knowledge ${\cal K}_0$ and channel ${\cal N}_0$. For applications with different background knowledge, we only need to redesign and train part of the semantic encoder and decoder layers and freeze the channel encoder and decoder layers. For different communication environments, we redesign and train part of the channel encoder and decoder layers and freeze the semantic encoder and decoder layers. If the knowledge and channel are totally different, the pre-trained transceiver can also reduce the time consumption because the weights of some layers in the pre-trained model can be reused in the new model even if the most layers need to redesign. After the other modules are trained, we will unfreeze them and train the whole network with few epochs to converge to the global optimum.

%\textcolor{black}{For example, when the transmitter and receiver are trained under channel A, with the deep TL approach, we only need to update several layers in the pre-trained transmitter and receiver instead of the whole network to make them work well under channel B. This results in much lower complexity.}

\section{Numerical Results}
In this section, we compare the proposed DeepSC with other DNN algorithms and the traditional source coding and channel coding approaches under the AWGN channels and Rayleigh fading channels, where we assume perfect CSI for all schemes. The transfer learning aided DeepSC is also verified under the erase channel and fading channel as well as different background knowledge. 

\subsection{Simulation Settings}
The adopted dataset is the proceedings of the European Parliament \cite{koehn2005europarl}, which consists of around 2.0 million sentences and 53 million words. The dataset is pre-processed into lengths of sentences with 4 to 30 words and is split into training data and testing data.

In the experiment, we set three Transformer encoder and decoder layer with 8 heads and the channel encoder and decoder are set as dense with 16 units and 128 units, respectively. For the mutual information estimation model, we set two dense layers with 256 units and one dense layer with 1 unit to mimic the function $T$ in \eqref{14}, where 256 units can extract full information and 1 unit can integrate information. These settings can be found in Table \ref{table 2}. For the baselines, we adopt joint source-channel coding based on neural network and the typical methods for separate source and channel codings.
\begin{itemize}
    \item DNN based joint source-channel coding \cite{gold2018}: The network consists of Bi-directional Long Short-Term Memory (BLSTM) layers. We label it as JSCC \cite{gold2018} in the simulation figures.
    \item Traditional methods: To perform the source and channel coding separately, we use the following technologies respectively:
    \begin{itemize}
        \item Source coding: Huffman coding, fixed-length coding (5-bit), and Brotli coding, where Brotli coding uses 2nd context model to compress the context information and every 128  sentences are compressed together in the simulation.
        \item Channel coding: Turbo coding \cite{heegard2013turbo} and Reed-Solomon (RS) coding \cite{reed1960polynomial}. We adopt turbo decoding method is log-MAP algorithm with 5 iterations. 
    \end{itemize}
\end{itemize} 
The BLEU and sentence similarity are used to measure the performance. The simulation is performed by the computer with Intel Core i7-9700 CPU@3.00GHz and NVIDIA GeForce GTX 2060.

\begin{figure*} 
\subfigure[AWGN]{
  \begin{minipage}[t]{1.0\linewidth} % 如果一行放2个图，用0.5，如果3个图，用0.33
    \centering 
    \includegraphics[width=160mm]{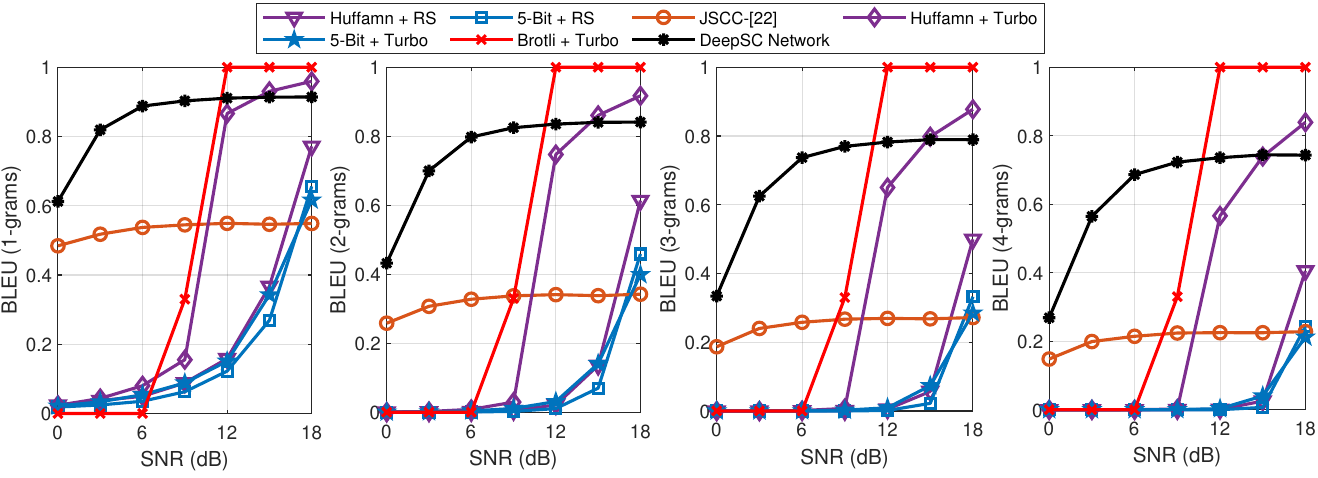}
    \label{grams} 
  \end{minipage}}

\subfigure[Rayleigh Fading]{
  \begin{minipage}[t]{1.0\linewidth} 
    \centering 
    \includegraphics[width=160mm]{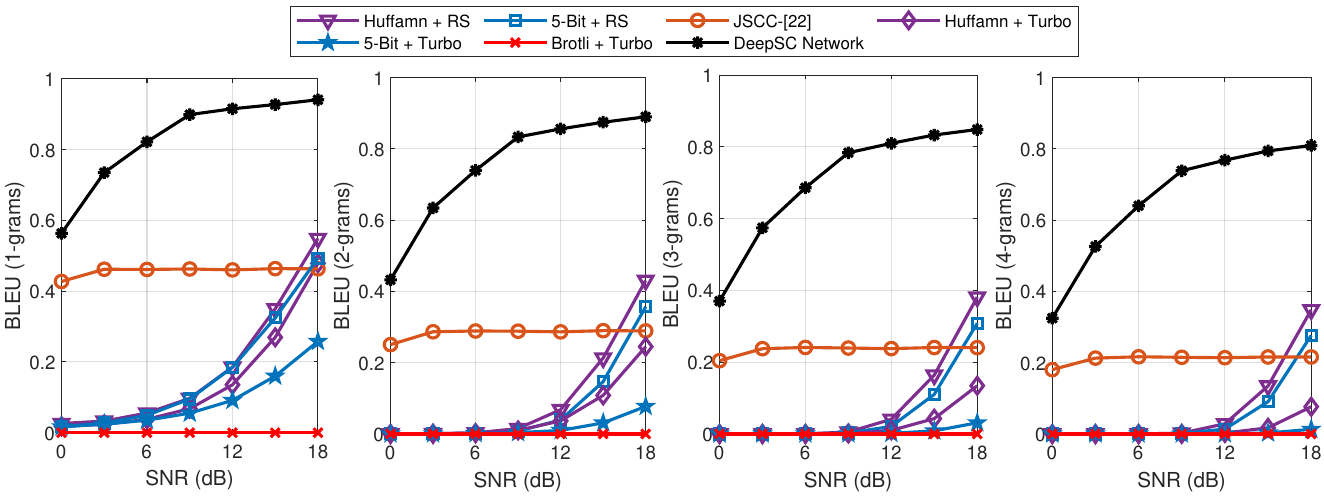}
    \label{rayleigh_grams} 
  \end{minipage} }
  
  \caption{BLEU score versus SNR for the same total number of transmitted symbols, with Huffman coding with RS (30,42) in 64-QAM, 5-bit coding with RS (42, 54) in 64-QAM, Huffman coding with Turbo coding in 64-QAM, 5-bit coding with Turbo coding in 128-QAM, Brotli coding with Turbo coding in 8-QAM; the DNN based JSCC \cite{gold2018} trained over the AWGN channels and Rayleigh fading channels, our proposed DeepSC trained over the AWGN channels and Rayleigh fading channels.}
  \label{bleu score}
\end{figure*}

\begin{figure*} 
\hspace{-3mm}
\subfigure[AWGN]{
  \begin{minipage}[t]{0.5\linewidth} % 如果一行放2个图，用0.5，如果3个图，用0.33
    \centering 
    \includegraphics[width=70mm]{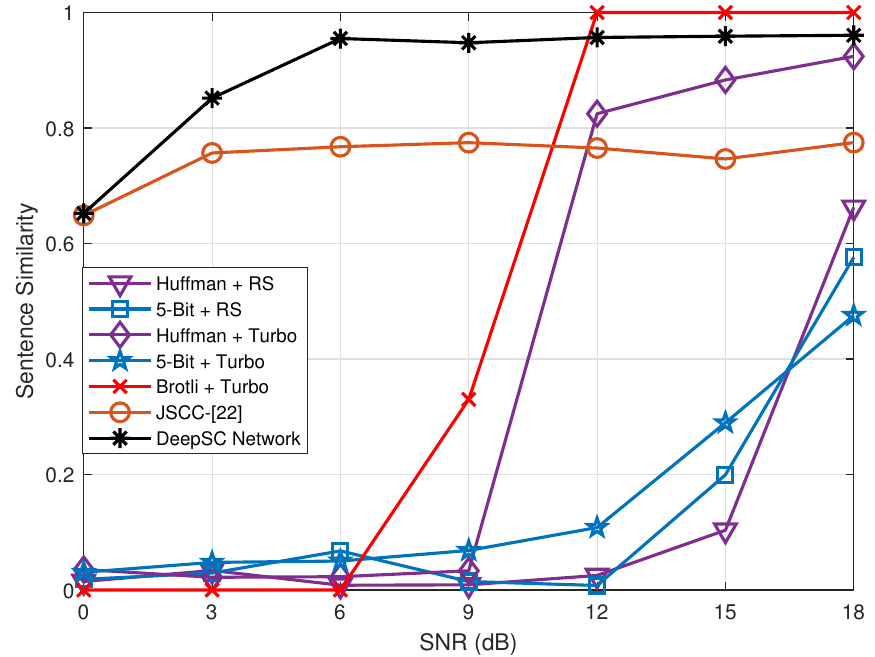}
    \label{sentecne similarity awgn} 
  \end{minipage}} \hspace{-1mm}
\subfigure[Rayleigh Fading]{
  \begin{minipage}[t]{0.5\linewidth} 
    \centering 
    \includegraphics[width=70mm]{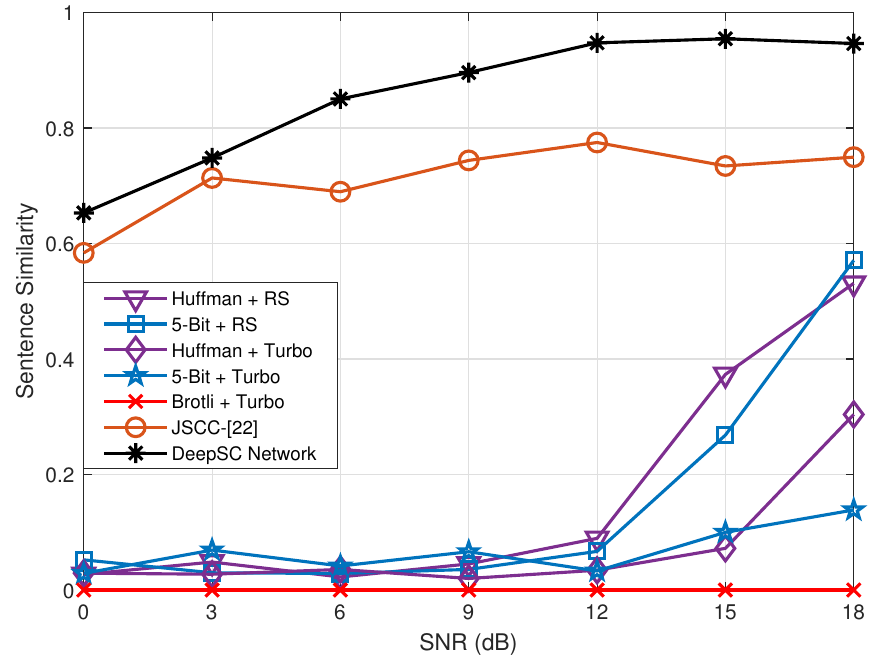}
    \label{sentecne similarity rayleigh} 
  \end{minipage} }
  
  \caption{Sentence similarity versus SNR for the same total number of transmitted symbols, with Huffman coding with RS (30,42) in 64-QAM; 5-bit coding with RS (42, 54) in 64-QAM; Huffman coding with Turbo coding in 64-QAM; 5-bit coding with Turbo coding in 128-QAM; Brotli coding with Turbo coding in 8-QAM; an E2E trained over the AWGN channels and Rayleigh fading channels~ \cite{gold2018}; our proposed DeepSC trained over the AWGN channels and Rayleigh fading channels.}
  \label{sentence similarity}
\end{figure*}

\subsection{Basic Model}

%这个地方画 constellations

Fig. \ref{bleu score} shows the relationship between the BLEU score and the SNR under the same  number of transmitted symbols over AWGN and Rayleigh fading channels, where the traditional approaches use 8-QAM, 64-QAM, and 128-QAM for the modulation.  Among the traditional baselines in Fig. \ref{grams}, Brotli coding outperforms the Huffman and fixed-length encoding over AWGN channels when the turbo coding is adopted for channel coding. The traditional approaches perform better than the DNN based method when the SNR is above 12 dB since the distortion from channel is decreased, where the Brotli with turbo coding performs better than the DeepSC. We observe that all DL enabled approaches are more competitive in the low SNR regime. 

%\textcolor{red}{Although the Huffman coding and Brotli coding with Turbo coding and the DeepSC have the similar performance in 1-gram when SNR is 12 dB or higher, the gap will increase from 1-gram to 4-grams, which means that the sentences decoded by the proposed approach can recover more semantic information than by the traditional approaches. }
\renewcommand\arraystretch{1.15}
\begin{table}[!t]
\footnotesize
\caption{The setting of the developed semantic network.}
\label{table 2}
\centering
\begin{tabular}{ |c| c |c |c |} 
\hline
& Layer Name&  Units &  Activation \\
\hline
\multirow{3}{4.5em}{\centering Transmitter\\(Encoder)} & 3$\times$Transformer Encoder & 128 (8 heads) & Linear \\ 
\cline{2-4}
& Dense & 256 & Relu \\
\cline{2-4}
& Dense & 16 & Relu \\
\hline
Channel & AWGN  & None & None \\
\hline
\multirow{4}{4.5em}{\centering Receiver\\(Decoder)} & Dense & 256 & Relu \\
\cline{2-4}
& Dense & 128 & Relu \\
\cline{2-4}
& 3$\times$Transformer Decoder & 128 (8 heads) & Linear \\
\cline{2-4}
& Prediction Layer & Dictionary Size & Softmax \\
\hline
\multirow{3}{4.5em}{\centering MI Model} & Dense & 256 & Relu \\
\cline{2-4}
& Dense & 256 & Relu \\
\cline{2-4}
& Dense & 1 & Relu \\
\hline
\end{tabular}
\end{table}

In Fig. \ref{rayleigh_grams}, the DL enabled approaches outperform all traditional approaches over  the Rayleigh fading channels, where RS coding is better than turbo coding in terms of 2-grams to 4-grams. This is because RS coding is linear block coding with long block-length, and can correct long series of bits, however, turbo coding is a type of convolutional coding with short block-length, so that the adjacent words have higher error rate. DeepSC is not only suitable for short block-length but also performs better in decoding adjacent words, i.e., 4-grams. Note that the BLEU score of the method with Brotil coding and turbo coding is always 0 over Rayleigh fading channels. This is because that 128 sentences are compressed together, while Brotil decoding requires error-free codes after channel decoding for the codes corresponding to the 128 sentences. However, it is almost to guarantee the error-free transmission over Rayleigh fading channels. Therefore, we fail to restore any of the 128 sentences compressed together in Brotil coding as shown in Fig. \ref{rayleigh_grams}. Besides, the lower BLEU score of the DL enabled approaches may not be caused by word errors. For example, it may be due to substitutions of words using synonyms or rephrasing, which does not change the meaning of the word. Fig. \ref{bleu score} also demonstrates that the joint semantic-channel coding design outperforms the traditional methods, which provides solution to \textit{Question 1} and \textit{3}.

\begin{table*}[htbp]
\small
\caption{The sample sentences between different methods over Rayleigh fading channels when SNR is 18 $dB$.}
\label{table 333}
\centering
\begin{tabular}{ |c| c |} 
\hline
Transmitted sentence & it is an important step towards equal rights for all passengers. \\
\hline
DeepSC &  it is an important step towards equal rights for all passengers.  \\
\hline
JSCC-[22] & it is an essential way towards our principles for democracy.  \\
\hline
Huffman + Turbo coding &  rt is a imeomant step  tomdrt equal  rights for atp passurerrs.  \\
\hline
Huffman + RS coding & it is an important step towards ewiral rlrsuo for all passengess.  \\
\hline
Bit5 + Turbo coding & it is an yoportbnt ssep sowart  euual qighd  fkr ill passeneers.  \\
\hline
Bit5 + RS coding & it iw an ymp!rdbnd stgo to!atds eq.al ryghts dkr alk passengers.  \\
\hline
\end{tabular}
\end{table*}

%Therefore, BLEU score is not a good measure for semantic information. Through extracting semantic information effectively, Fig. \ref{grams} also demonstrates that the joint semantic-channel coding design outperforms the traditional one, which provides solution to \textit{Question 1} and \textit{3}.

% 这个地方画 SNR with Similarity (这幅图得加traditional 对比）
% syntax
Fig. \ref{sentence similarity} shows that the proposed performance metric, the sentence similarity, with respect to the SNR under the same total number of symbols, where the traditional approaches use 8-QAM, 64-QAM and 128-QAM. In Fig. \ref{sentecne similarity awgn}, the proposed metric has shown the same tendency compared with the BLEU scores. Note that for part of the traditional methods, i.e., Huffman with Turbo coding, even if it can achieve about 20\% word accuracy in BLEU score (1-gram) from Fig. \ref{grams} when SNR = 9 dB, people are usually unable to understand the meaning of  texts full of errors. Thus, the  sentence similarity in Fig. \ref{sentecne similarity awgn} almost converges to 0. For the  DeepSC, it achieves more than 90\% word accuracy in BLEU score (1-gram) when SNR is higher than 6 dB in Fig. \ref{grams}, which means people can  understand the texts well. Therefore the  sentence similarity tends to 1.  Fig. \ref{rayleigh_grams} and Fig. \ref{sentecne similarity rayleigh} show the same tendency. The benchmark, including the DNN based JSCC method in \cite{gold2018} under  Rayleigh fading channels, also gets much higher score than the traditional approaches in terms of the sentence similarity since it can capture the features of the syntax and the relationship of the words, as well as present texts that is easier for people to understand. Few representative results are shown in Table \ref{table 333}.

In brief, we can conclude that the tendency in sentence similarity is more closer to  human judgment and the DeepSC achieves the best performance in terms of both BLEU score and sentence similarity. Compared to the simulation results with BLEU score as the metric, the sentence similarity score can better measure the semantic error, which solves the \textit{Question 2}.

\begin{figure}[t!]
	\centering
	\includegraphics[width=70mm]{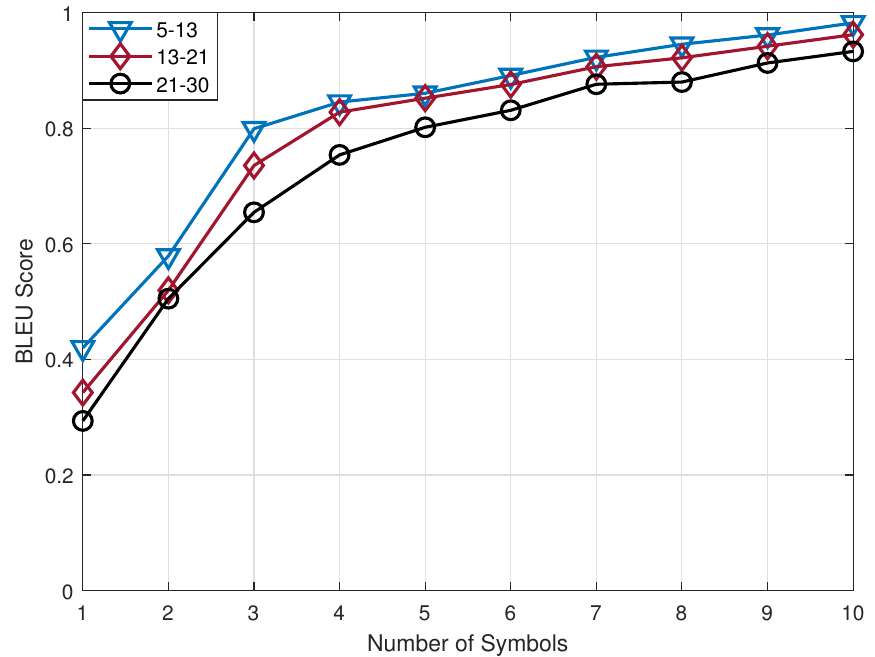}
	\caption{BLEU score (1-gram) versus the average number of symbols used for one word in the DeepSC, SNR = 12 dB.}
	\label{symbol_with_wer}
\end{figure}

Fig. \ref{symbol_with_wer} illustrates that the impact of the number of symbols per word on the 1-gram BLEU score when SNR is 12 dB. \textcolor{black}{As} the number of symbols per word grows, the BLEU scores increase significantly due to the increasing distance between constellations gradually. Generally, people can  understand the basic meaning of transmitted sentences with over 85\% word accuracy in BLEU score (1-gram). For short sentences consisted \textcolor{black}{of} 5 to 13 words, our proposed DeepSC can achieve 85\% accuracy with 4 symbols per word, which means that we can use fewer symbols to represent one word in the environment that mainly transmits short sentences. Therefore, it can achieve high speed transmission rate.  For longer sentences consisted from \textcolor{black}{of} 21 to 30 words, the proposed DeepSC faces more difficulties to understand the complex structure of the sentences in \textcolor{black}{the} transmitted texts. Hence the performance is degraded with longer sentences. One way to improve the BLEU score is to increase the average number of symbols used for each word.

\begin{figure}[t!]
	\centering
	\includegraphics[width=70mm]{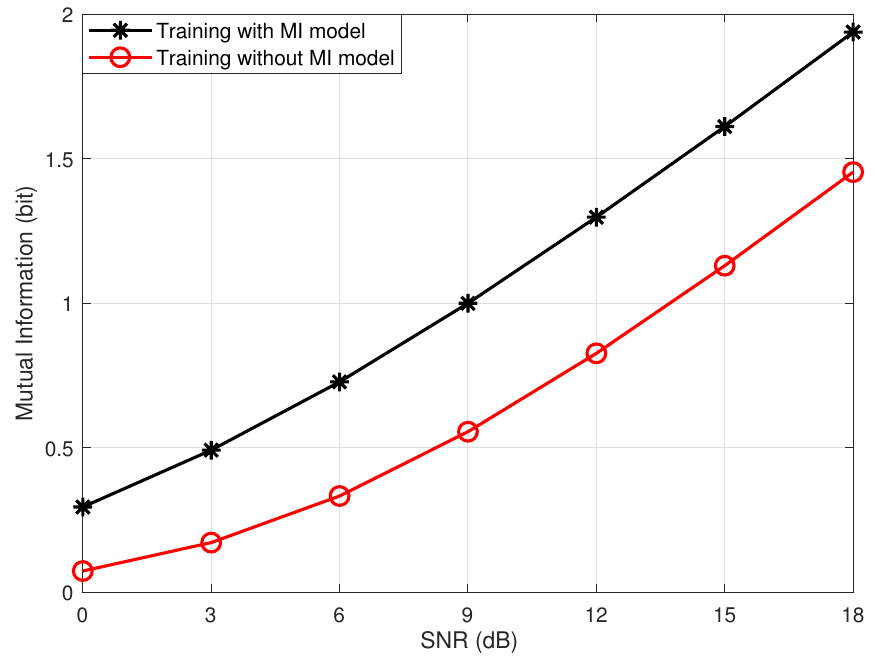}
	\caption{SNR versus mutual information for different trained encoders, with 8 symbols per word.}
	\label{snr_mi}
\end{figure}

%\textcolor{black}{However, as shown in Fig. \ref{tl_epochs} (b), we can note that the convergence speed of the DeepSC under the fading channel is . This is because that the fading channel features are modelled by a mathematical model rather than  a neural network, so that it cannot provide the complete backward gradient during training.}

\subsection{Mutual Information}

Fig. \ref{snr_mi} demonstrates the relationship between SNR and mutual information after training. \textcolor{black}{As we can imagine}, the mutual information increases \textcolor{black}{with SNR}. From the figure, the performance of the transceiver trained with the mutual information estimation model outperforms that without such a model. From Fig. \ref{snr_mi}, with the proposed mutual information estimation model, the obtained mutual information at SNR = 4 dB is approximately same as that without the training model at SNR = 9dB. From another point of view, the mutual information estimation model leads to better learning results, i.e., data distribution, at the encoder to achieve higher data rate. In addition, this shows that introducing  \eqref{eq14} in loss function can improve the mutual information of the system.

\begin{figure}[t!]
	\centering
	\includegraphics[width=70mm,height=55mm]{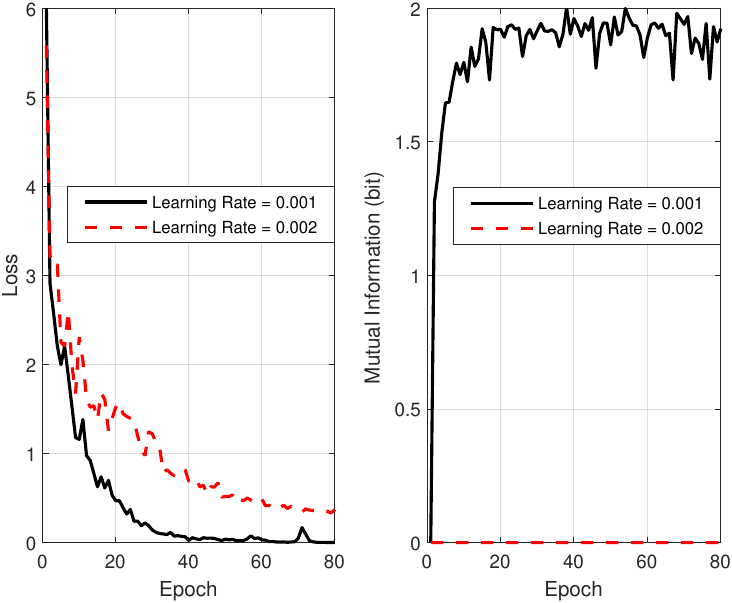}
	\caption{The impact of different learning rates with training SNR = 12 dB.}
	\label{loss_mutual}
\end{figure}

%\begin{figure}[t!]
%	\centering
%	\includegraphics[width=80mm, height = 60mm]{picture//simulation//MI//SNR_BLEU_MI.eps}
%	\caption{BLEU score (1-gram) versus SNR for different learning rates, with training SNR = 12 dB.}
%	\label{loss_mutual_2}
%\end{figure}

\begin{figure*} 
  \begin{minipage}[t]{0.49\linewidth} % 如果一行放2个图，用0.5，如果3个图，用0.33
    \centering 
    \includegraphics[width=70mm]{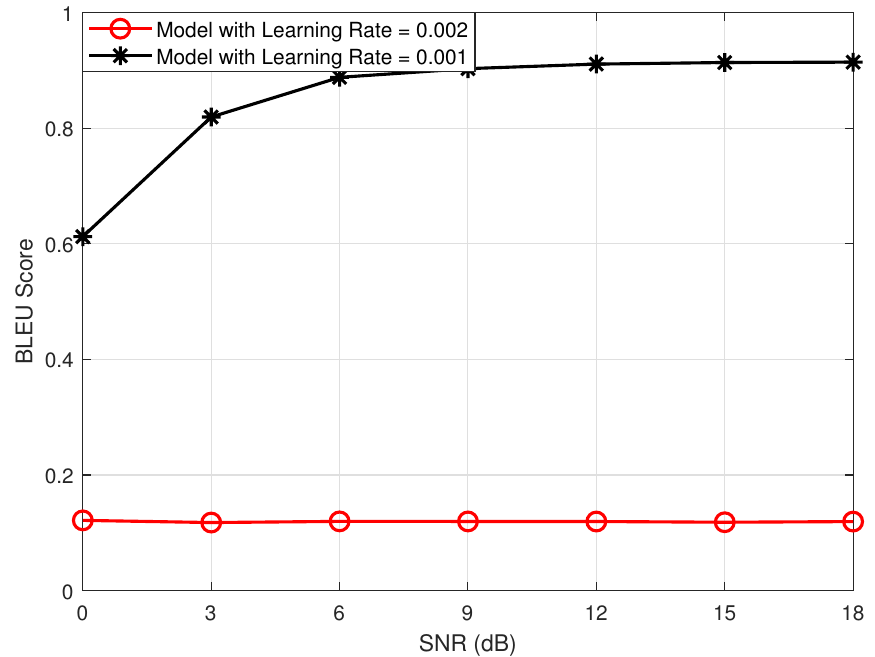} 
    \caption{BLEU score (1-gram) versus SNR for different learning rates, with training SNR = 12 dB.} 
    \label{loss_mutual_2} 
  \end{minipage}
  \hspace{3mm}
  \begin{minipage}[t]{0.49\linewidth} 
    \centering 
    \includegraphics[width=70mm,height=55mm ]{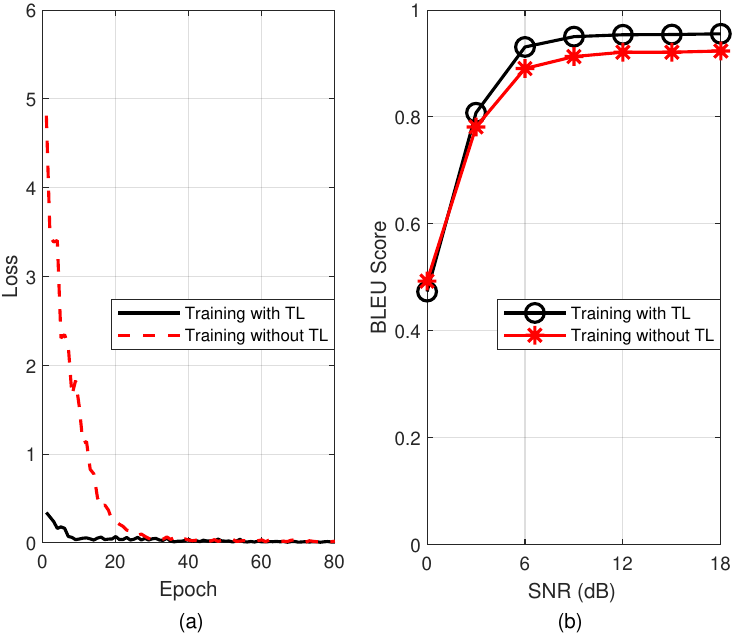} 
    \caption{Transfer learning (TL) aided DeepSC with different background knowledge: (a) loss values versus the number of training epochs, (b) BLEU score (1-gram) versus the SNR.} 
    \label{tl_performance} 
  \end{minipage} 
\end{figure*}

Fig. \ref{loss_mutual} draws the \textcolor{black}{relationship} between the loss value in \eqref{total loss} and the mutual information with increasing epoch. Fig. \ref{loss_mutual_2} indicates the relationship between BLEU score and SNR. The two figures are based on models with the same structure but different training parameters, i.e., learning rate. In Fig. \ref{loss_mutual}, the obtained mutual information is different, i.e., the mutual information of model with learning rate 0.001 increases along with decreasing loss value while the other one with learning rate 0.002 stays zero although the loss values of two models gradually converge to a stable state. From Fig. \ref{loss_mutual_2}, the BLEU score with learning rate 0.001 outperforms that with learning rate 0.002, which means that even if the neural network converges to \textcolor{black}{a} stable state, it is possible that gradient decreases to \textcolor{black}{a} local minimum instead of the global minimum. During the training process, the mutual information can be used as a tool to decide whether the model converges effectively.

\subsection{Transfer Learning for Dynamic Environment}
In this experiment, we present the performance of transfer learning aided DeepSC for two tasks: transmitter and receiver re-training over different channels and diffident background knowledge. 

%\begin{figure}[t]
%	\centering
%	\includegraphics[width=80mm, height = 60mm]{picture//simulation//Transfer_Learning//TL_different_knowledge.pdf}
%	\caption{Transfer learning (TL) aided DeepSC with different knowledge: (a) loss values versus the number of training epochs, (b) BLEU score (1-gram) versus the SNR.}
%	\label{tl_performance}
%\end{figure}

\begin{figure*}[htb]
	\centering
	\includegraphics[width=140mm,height=60mm]{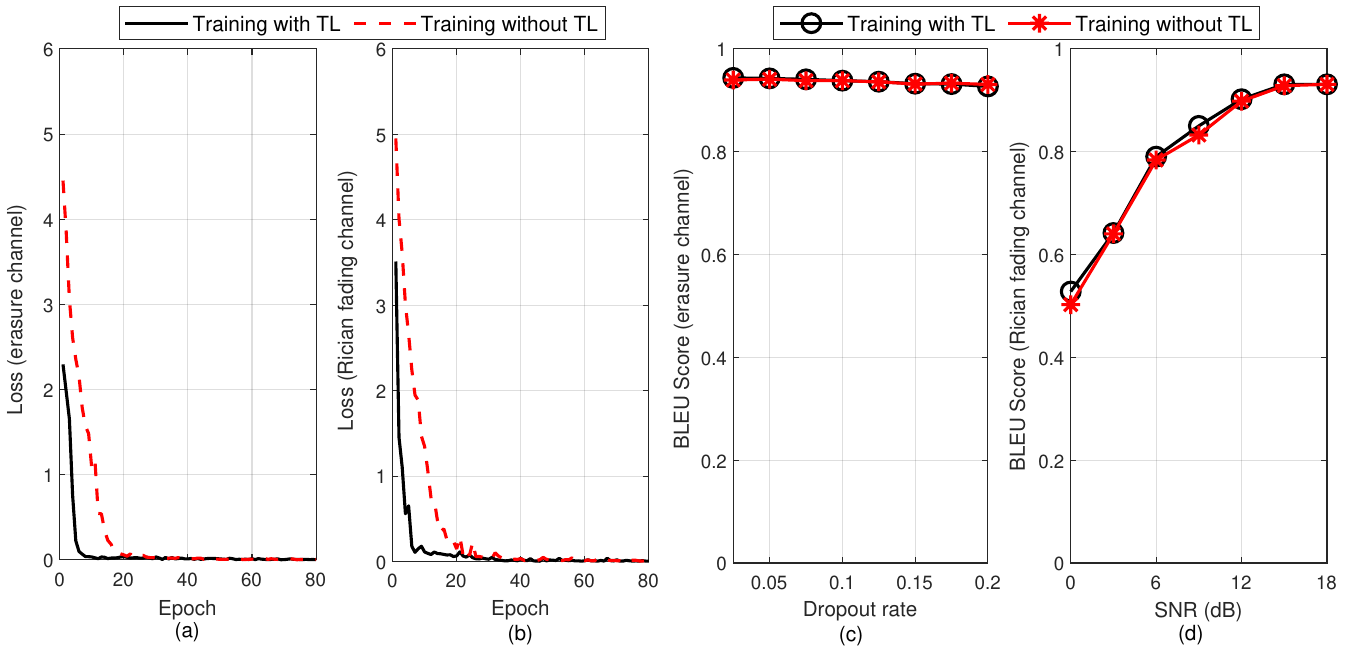}
	\caption{Transfer learning aided DeepSC with different channels: (a) loss values versus  epochs under the  erasure channel; (b) Loss values versus epochs under the Rician fading channel; (c) BLEU score (1-gram) versus the dropout rate;  (d)  BLEU score (1-gram) versus the SNR.}
	\label{tl_epochs}
\end{figure*}

%这个地方画 SNR with Similarity
Fig. \ref{tl_performance} shows the training efficiency and the performance for different background knowledge, where the model will be trained and re-trained in new background knowledge with the same channel (AWGN) for different background knowledge. The models have the same structure and re-train with the same parameters in each scenario. From Fig. \ref{tl_performance}(a), the epochs are reduced from 30 to 5 to reach convergence. In Fig. \ref{tl_performance}(b), the pre-trained model can provide additional knowledge so that the corresponding model training outperforms that of re-training the whole system. This demonstrates that the transfer learning aided DeepSC can help the transceiver to accommodate the new requirements of communication environment.

Fig. \ref{tl_epochs}  shows the training efficiency and the performance for different channels, where the DeepSC transceiver is pre-trained under the AWGAN channel, and then it is re-trained under the erasure channel and the Rician fading channel, respectively, with the same background knowledge. The models have the same structure and re-train with the same parameters in each scenario. From Fig. \ref{tl_epochs}(a) and Fig. \ref{tl_epochs}(b), the adoption of the pre-trained model can speed up the training process for both the erasure channel and Rician fading channel.  In Fig.~\ref{tl_epochs}(c) and Fig.~\ref{tl_epochs}(d), the performance of the DeepSC with pre-trained model is similar to that  without pre-trained model  channel while the required complexity is  reduced significantly as less number of epochs is required during the re-training process.   It is further noted that the  BLEU score achieved by the DeepSC is slightly degraded under the fading channel, especially in the lower SNR region, compared to that under the erasure channel. 

\subsection{Complexity Analysis}

The computational complexities of the proposed DeepSC, the JSCC in \cite{gold2018}, the RS coding, Turbo coding, are compared in Table \ref{table-3} in terms of the average processing runtime per sentence\footnote{The runtime of source coding and decoding are omitted in the comparison.}.  All the DL enabled approaches have lower runtime than the traditional approaches, where turbo coding costs much longer runtime in log-map iterations and  the JSCC  \cite{gold2018} requires the lowest average time due to its simple network architecture, however, it comes with poorer semantic processing capability. As a comparison, the runtime of our proposed DeepSC significantly outperforms the traditional schemes and is slight higher than JSCC \cite{gold2018} but with significant performance improvement.
\begin{table}[htbp]
\caption{The average sentence processing runtime versus various schemes. }
\label{table-3}
\centering
\begin{tabular}{ |c| c |c |c|c|} 
\hline
& DeepSC &  JSCC [22] &  RS coding & Turbo coding\\
\hline
Runtime  & 3.27ms&  2.71ms &  4.14ms  & 8.59ms\\
\hline
\end{tabular}
\end{table}

\section{Conclusions}
In this paper, we have proposed a semantic communication system, named DeepSC, which jointly performs the semantic-channel coding for texts transmission.  With the DeepSC,  the length of input texts and output symbols are variable, and the mutual information is considered as a part of the loss function to achieve higher data rate. Besides, the deep transfer learning  has been adopted to meet different transmission conditions and speed up the training of new networks by exploiting the knowledge from the pre-trained model. Moreover, we initialized sentence similarity as a new performance metric for the semantic error, which is a measure closer to human judgement. The simulation results has demonstrated that the DeepSC  outperforms  various benchmarks, especially in the low SNR regime. The proposed transfer learning aided DeepSC has shown its ability to adapt to different channels and knowledge with fast convergence speed. Therefore, our proposed DeepSC is a good candidate for text transmission, especially in the low SNR regime, which could be very useful for cases with massive number of devices to be connected with the limited spectrum resource. 

We conclude the difference between semantic communication systems  and conventional communication systems into the following:
\begin{enumerate}
    \item Different data processing domains. The former  process data in semantic domain while the latter compress data in entropy domain.
    \item Different communication targets. The conventional communication systems focus on the exact data recovery while the semantic communication systems serve for the decisions or targets of the transmission.
    \item Different system designs. The conventional systems only design and optimize the information transmission modules, which are contained in the traditional transceiver, however, the semantic systems jointly design the whole information processing blocks from source information to final  targets of applications.
\end{enumerate}  
Following the concept of semantic communications proposed in this paper, we have developed L-DeepSC \cite{XieQ21} and DeepSC-S \cite{WengSpeech} for text  and speech transmission.

% or


%\appendices


%\section*{Acknowledgment}


\ifCLASSOPTIONcaptionsoff
  \newpage
\fi



\bibliographystyle{IEEEtran}
\bibliography{reference.bib}
\end{document}